\documentstyle[aps,preprint,epsf]{revtex}

\begin{document}
\preprint{\sl Submitted to Phys. Rev. Letters}

\title{The Effect of Strain on Thermodynamics of the Weakly
First-Order Phase Transition}

\author {M.A.Fradkin\cite{IKAN}}

\address{Department of Mechanical and Aerospace Engineering  \\
                Carleton University, Ottawa, Ont. K1S 5B6, Canada \\
                {\em e-mail:} {\tt mfradkin@next.mrco.carleton.ca}}

\draft
\maketitle

\begin{abstract}
Elastic matrix distortion around a growing inclusion of a new phase is analyzed
and the associated contribution to the Gibbs free energy is considered. The
constant-composition transformation from the parent to product phase is
considered within the frame of Landau theory of phase transitions. The volume
misfit between the inclusion and matrix is assumed to originate from the
transformation volume change coupled with the phenomenological order parameter.
The minimization of free energy with respect to the volume change and order
parameter gives the dependence of Gibbs energy on the volume fraction of the
product phase. The transformation proceeds in a finite temperature region with
the equilibrium volume fraction dependent on temperature rather than at a fixed
temperature as it would be expected for the first-order transition. The
activation processes are shown to be irrelevant and the transformation kinetics
is found to be fluctuationless.
\end{abstract}


\narrowtext

Spontaneous strain is known to play an important role in many cases of the
phase transformations in solids\cite{Christian,KhachBook}. Due to the
difference in crystalline structures and elastic properties of the co-existing
phases the volume and shape misfits appear between the inclusion of the product
phase and surrounding matrix of the parent one. An associated elastic strain
both in the inclusion and in the matrix tends to relax this misfit by the cost
of an additional elastic energy\cite{Eshelby}. The minimization of this elastic
energy defines the shape and crystallographic orientation of the new phase
particles\cite{Royt}. The elastic effects change qualitatively the phase
diagram of heterophase systems and the phase rule is no longer valid since it
does not take into account the elastic interaction between the phases which
co-exist in a state of coherent equilibrium\cite{LarcheCahn}. The `chemical'
free energy alone can not provide the equilibrium fractions of the phases with
different composition through well-known double-tangent
construction\cite{CahnLarche84,JohnsonVoorees}.

The elastic effects are considered in the present paper for the case of the
first-order solid-state phase transition without change in a chemical
composition. Such a transition may take place in pure elements,
fixed-composition compounds and in alloys with very slow diffusion kinetics.
The stresses are shown to result in a finite  temperature interval of the phase
co-existence with the equilibrium phase fractions depending on temperature
whereas this co-existence appears at the single temperature point in the
absence of elastic effects. The transition kinetics is shown to be changed as
well and the activation processes does not control the transition development.

A usual approach\cite{Eshelby} to elastic effects is to let the transformation
proceed in the inclusion disconnected from the matrix, to let the `zero-stress'
transformation strain tensor to appear and than to bring the inclusion into an
elastic contact with the surrounding matrix. The strain energy is then
calculated using disconnected inclusion as an elastic reference state for the
product phase. Though the new phase is sometimes considered to have different
elastic moduli, calculations are mostly being done in the linear elasticity
theory. However, when the transformation is sensitive to an applied external
stress, a `zero-stress' assumption is no longer valid and a non-linear
dependence of the free energy of the new phase on the stress level should be
considered.

The simplest possible way to analyze the transition strain is to use Landau
theory of phase transitions\cite{LandLif,Toledano} taking into account the
coupling of the phenomenological order parameter with the strain tensor
components. Such a coupling has been recently considered for the case of a
proper ferroelastic (martensitic) phase transition\cite{me} and both the effect
of hydrostatic and uni-axial pressure have been studied. In the present paper I
consider the elastic interaction in a heterophase system for a model case of
the spherical inclusions of the product phase embedded into an isotropic
elastic matrix of the parent phase. The quadratic coupling of the order
parameter with the volume change that corresponds to a variety of systems ({\em
e.g.} see\cite{mefull}) is analysed.

The remaining part of this paper is organized as follows. We begin with a
linear elasticity analysis of the displacement field and the elastic energy
associated with the volume change in an inclusion embedded into the isotropic
matrix. Then the free energy difference between the infinite crystals of parent
and product phases is considered within Landau theory and the coupling of the
phenomenological order parameter with elastic strain is taken into account. The
energy cost of the formation of a finite fraction of the new phase as
inclusions inside the bulk crystal of the parent phase is than considered.
After minimization of this free energy with respect to a volumetric strain, the
resulting expression that depends on the volume fraction of the new phase is
analyzed. Then the transformation kinetics is briefly discussed and the role of
fluctuations is analyzed.

Let us start with the spherical inclusion in an isotropic matrix and consider
the bulk crystal of the parent phase as a reference state for the elastic
energy calculation. A spherical shape of the new phase inclusion presupposes
either that transformation strain does not have shear components or that such
components disappear upon averaging over the ensemble of twinned domains of the
new phase composing the inclusion. The latter situation takes place in many
cases of martensitic phase transformations in metals.

The radial displacement field ${\bf u} (r)$ appears due to the transformation
and it has the following form determined by the elastic equilibrium
conditions\cite{LanLifEl}
\begin{displaymath}
{\bf u}(r) = {a_1 r \ \ \ \ {\rm for}  \ \ \ \ r \le R \atopwithdelims\{.
 {a_2 r + {b_2 \over {r^2}}} \ \ \ \ {\rm for} \ \ \ \  R < r \le R_0} \ .
\end{displaymath}
Here $R$ is the inclusion radius and $R_o = R/\sqrt[3]\nu$ is the radius of the
spherical domain of the parent phase attributed to the inclusion for $\nu$
being the volume fraction of new phase. The first boundary condition states
that the displacement field vanishes at $R_o$ and the continuity of the
displacement field across the interface implies the second boundary condition.
Hence, one can get
\begin{displaymath}
a_1 = {\epsilon_0 \over 3} \ , \ \ \
a_2 = {\epsilon_0 \over 3} \, {\nu \over {\nu - 1}} \ \ \ {\rm and}  \ \ \ \
b_2 = {\epsilon_0 \over 3} \, {R^3 \over {1 - \nu}} \ ,
\end{displaymath}
where $\epsilon_0 = {\rm Tr} (\hat{\epsilon})$ is a volumetric strain inside
the inclusion.

The elastic energy associated with the matrix strain and expressed per unit
volume of the {\em new} phase is given by an expression
\begin{displaymath}
\Delta {\cal G}_{el}^{(m)} = {K_0 \over 2} \, \epsilon_0^2 \,
	{\nu + \gamma \over {1 - \nu}} \ ,
\end{displaymath}
where $\gamma = (4\mu_0)/(3 K_0)$ is the renormed ratio between the shear and
bulk moduli of the parent phase. This energy corresponds to the stress that
plays a role of the external pressure applied to the inclusion of the product
phase.

The elastic energy (per unit volume of the new phase) associated with
volumetric strain inside the inclusion is
\begin{displaymath}
\Delta {\cal G}_{el}^{(i)} = {K_0 \over 2} \, \epsilon_0^2 \ .
\end{displaymath}

Total elastic energy that has to be added to the Gibbs energy difference
between the phases is equal to
\begin{eqnarray}
\label{elenrg}
\Delta {\cal G}_{el} (\epsilon_0) & = & \Delta {\cal G}_{el}^{(i)} + \Delta
{\cal G}_{el}^{(m)} \nonumber \\
& = & {K_0 \over 2} \, \epsilon_0^2 \, {1 + \gamma \over {1 - \nu}} \ .
\end{eqnarray}
This expression is valid for a sufficiently small $\nu$ when the inclusions of
the product phase are well separated and the overlapping of associated
spherical domains of the parent phase can be neglected.

The free energy difference between bulk crystals of the parent and product
phases which are related by the symmetry breaking phase transition can be
expanded in the mean-field Landau theory\cite{LandLif} in powers of the `order
parameter' $\eta$. The theory initially was developed for the second-order
transition where $\eta$ is continuous at the transition point, however,
weakly-discontinuous first-order transitions can be considered as
well\cite{Toledano}. If the symmetry groups of both parent and product phases
are known {\em a priory} then the scalar order parameter can be used and the
Ginzburg-Landau expansion of the free energy difference per unit of volume has
the form\cite{LandLif,Toledano}
\begin{equation}
\Delta {\cal G}_{\rm GL} (T,\eta) = {\alpha \over 2}(T - T_c)\eta^2 + {B \over
3}\eta^3  + {C \over 4}\eta^4 \ .
\label{firstfield}
\end{equation}
where $T_c$ is a critical temperature. Only the second-degree coefficient is
supposed to depend on temperature and the equilibrium value of $\eta$ is
determined by the minimization of $\Delta {\cal G}$. Stability requires the
highest order coefficient $C$ to be positive and the third degree term $B \neq
0$ implies the first order of the transition.

Gibbs free energy (\ref{firstfield}) has two possible minima. One with $\eta =
0$ corresponds to a high-temperature undistorted phase stable for $T \ge T_c$.
For
$$T \le T_0 = T_c + {1 \over 4}{B^2 \over {\alpha C}}$$
a low-symmetry phase exists with
\begin{equation}
        \eta = - {B \over {2 C}} \left(1 + \left({T_0 - T \over {T_0 - T_c}}
\right)^{\frac{1}{2}} \,\right) . \label{FOE}
\end{equation}
The phase energies become equal at the temperature of the first-order
transition $T_\ast = T_c + {2 \over 9}{B^2 \over {\alpha C}}$ where the
order parameter jumps from the $\eta = 0$ to \[ \eta = - {2 \over 3}\,{B \over
C}, \] overcoming the activation energy barrier \[ \Delta {\cal G}_b = {1 \over
324}\,{B^4 \over C^3}. \]

To analyze an associated strain effect the coupling of $\eta$ with the strain
tensor $\hat{\epsilon}$ has to be considered and the Ginzburg-Landau expansion
should include the elastic terms (\ref{elenrg}). In the simplest case the
symmetry allows the volumetric strain $\epsilon_0 = {\rm Tr} (\hat{\epsilon})$
to be coupled with $\eta$ in the lowest order by the following term:
\begin{equation}
        \Delta {\cal G}_{int} (\epsilon_0, \eta) = D_0 \epsilon_0 \eta^2 \ .
        \label{inter}
\end{equation}
Adding this term to the elastic energy (\ref{elenrg}) and finding the minimum
of the resulting expression with respect to $\epsilon_0$ we obtain the
dependence of the volumetric strain inside the inclusion on the
phenomenological order parameter in the form
\begin{equation}
\epsilon_0 (\eta) = - {D_0 \over K_0} \, \eta^2 \, {{1-\nu} \over {1 + \gamma}}
\end{equation}
This leads to a renormed Ginzburg-Landau expansion of $\Delta {\cal G}$ in
powers of $\eta$\cite{me}:
\begin{eqnarray}
\Delta {\cal G}(T,\eta) & = & \Delta {\cal G}_{\rm GL} (T,\eta) + \Delta {\cal
G}_{el} (\epsilon_0 (\eta)) + \Delta {\cal G}_{int} (\epsilon_0(\eta), \eta)
\nonumber \\ & = &  {\alpha \over 2}\,(T - T_c)\,\eta^2 + {B \over 3}\, \eta^3
\nonumber \\ & &
+  {C \over 4} \, \eta^4 \left( 1 - {2 D_0^2 \over {K_0 C}} \, {{1-\nu} \over
{1 + \gamma}}\right) \ .
\label{RTOT}
\end{eqnarray}

\begin{figure}
\epsfxsize 5.75in
\centerline{\epsffile{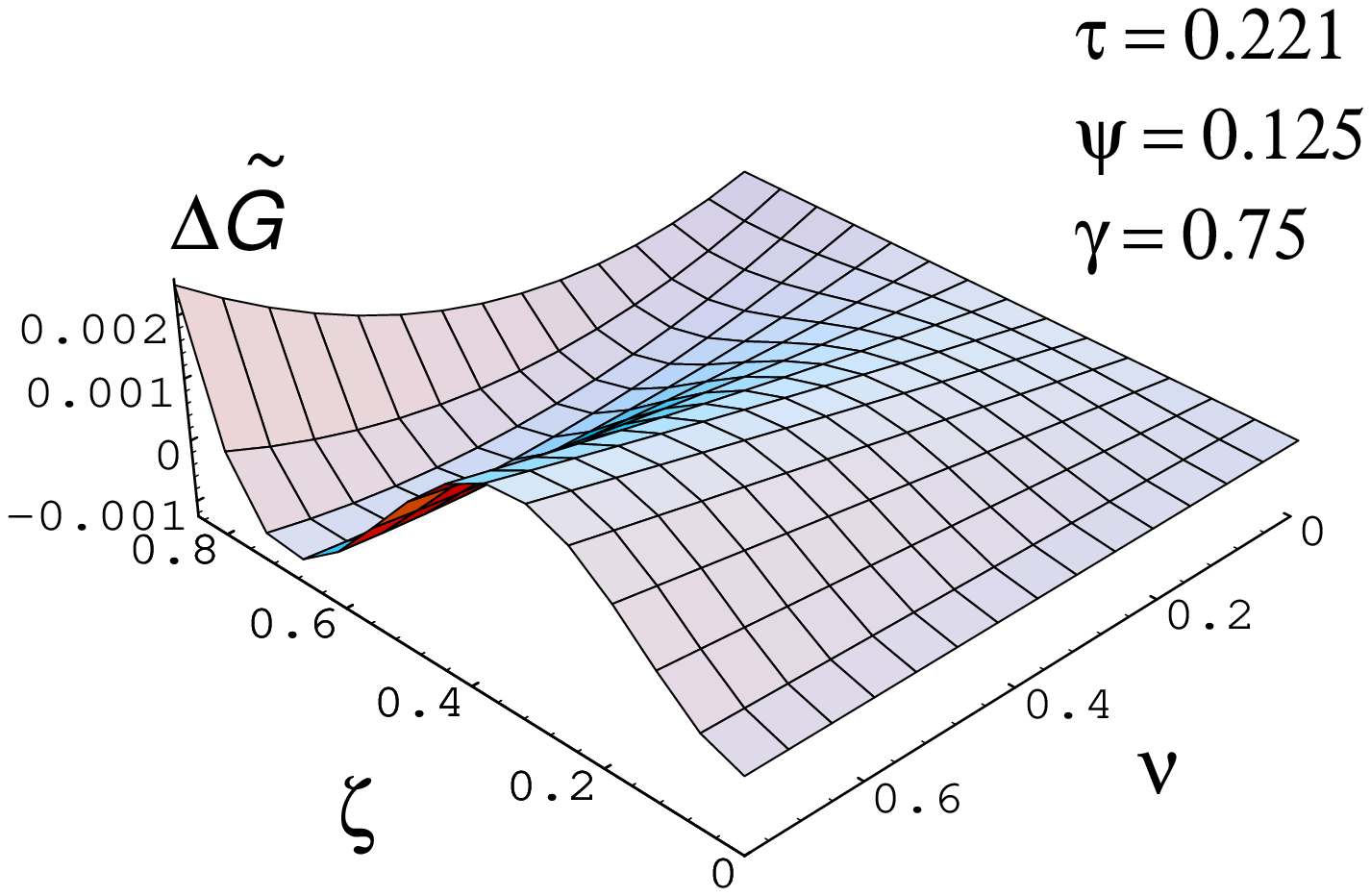}}
\caption{Dependence of $\Delta \tilde{\cal G}$ on the order parameter $\zeta$
and the volume fraction of the new phase $\nu$.}
        \label{Energy3D}
\end{figure}

We have considered the unstrained parent phase as a reference state for an
energy calculations, thus, the energy cost for the formation of the new phase
(\ref{RTOT}) should be multiplied by its volume fraction. Choosing the case of
$B < 0$ which implies a positive $\eta$ in the product phase, we can write the
expansion of total free energy per unit {\em system} volume in the following
form
\begin{eqnarray}
\Delta \tilde{\cal G} & = & {\frac{C^3}{B^4}} \Delta {\cal G} \nonumber \\ & =
&
\nu \left( {\tau \over 2} \ \zeta^2 - {\zeta^3 \over 3} + {\zeta^4 \over 4}
\left( 1  - {{\psi (1-\nu)} \over {1 + \gamma}}\right) \right)\ ,
	\label{rescaled}
\end{eqnarray}
with \[ \eta = - {B \over C} \,\zeta \ , \ \
        \tau = {{\alpha C} \over B^2} \ (T - T_c) \  \
        {\rm and} \ \ \psi = {2 D_0^2 \over {K_0 C}} \ . \]
The dependence of $\Delta \tilde{\cal G}$ on $\zeta$ and $\nu$ is shown in
Fig.\ref{Energy3D} for some $\tau$, $\psi$ and $\gamma$.

\begin{figure}
\epsfxsize 5.75in
\centerline{\epsffile{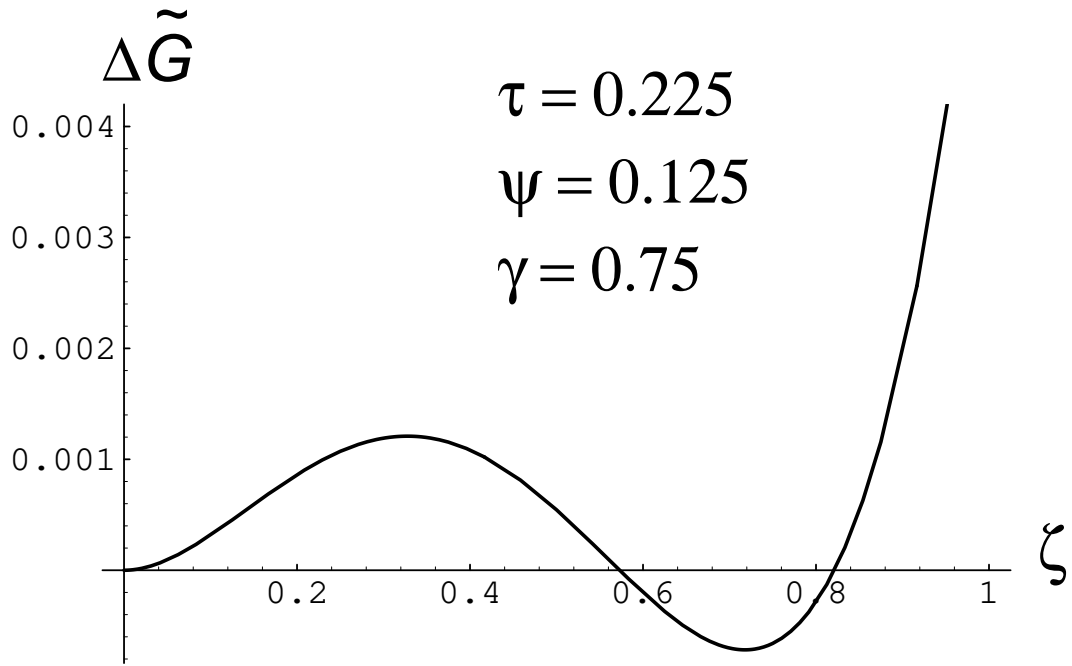}}
\caption{Free energy {\em vs} the phenomenological order parameter $\zeta$.}
        \label{EnergyZeta}
\end{figure}

\begin{figure}
\epsfxsize 5.75in
\centerline{\epsffile{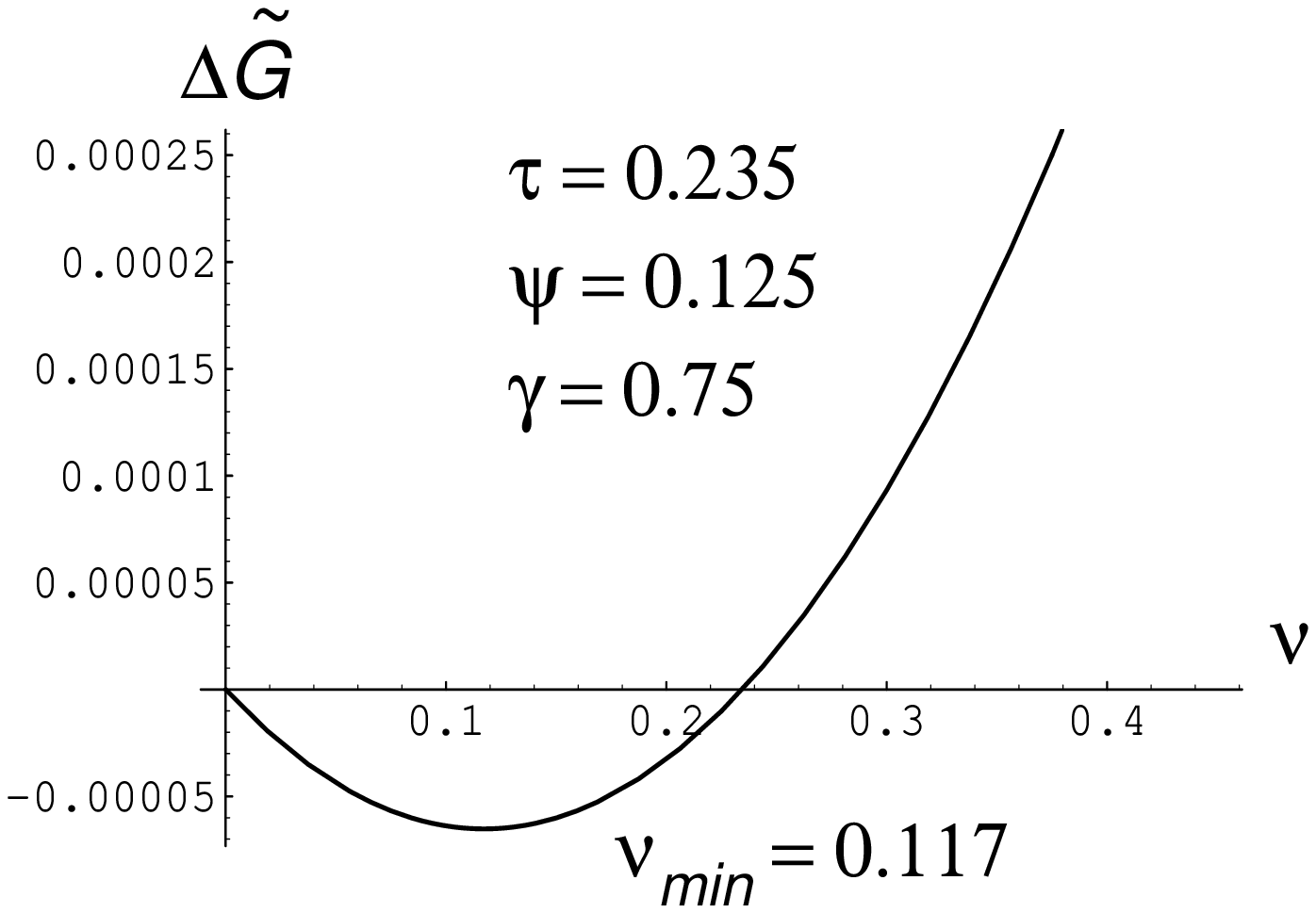}}
\caption{Dependence of $\Delta \tilde{\cal G}$ on the volume fraction of the
new phase $\nu$. Equilibrium $\nu$ for these $\tau$, $\psi$ and $\gamma$ is
0.117.}
        \label{EnergyNu}
\end{figure}

The state of equilibrium is determined by the minimum of this free energy with
respect to both $\zeta$ and $\nu$
\begin{mathletters}
\label{minimum}
\begin{eqnarray}
\frac{\partial \Delta \tilde{\cal G}}{\partial \zeta} = 0  \ \ \ \ \ {\rm and}\
\ \ \  \ \frac{\partial \Delta \tilde{\cal G}}{\partial \nu} = 0 	\label{equal}
\\
\frac{\partial^2 \Delta \tilde{\cal G}}{\partial^2 \zeta}  +
\frac{\partial^2 \Delta \tilde{\cal G}}{\partial^2 \nu} > 0 \label{cond1} \\
\frac{\partial^2 \Delta \tilde{\cal G}}{\partial^2 \zeta} \, \frac{\partial^2
\Delta \tilde{\cal G}}{\partial^2 \nu} - \left(\frac{\partial^2 \Delta
\tilde{\cal G}}{\partial \nu \partial \zeta}\right)^2 > 0 \ . \label{cond2}
\end{eqnarray}
\end{mathletters}
Solving coupled equations (\ref{equal}) we obtain
\begin{eqnarray}
\zeta & = & {2 \over 3} \, \left( 1 + {\psi \over {1 + \gamma - \psi}} \right)
\\
\nu & = & {{1 + \gamma - \psi}\over {2\,\psi }} -
  {{9\,{{\left( 1 + \gamma - \psi \right)}^2}\,\tau} \over
    {4\,\left( 1 + \gamma \right) \,\psi }} \ . \label{nutau}
\end{eqnarray}
The stability conditions (\ref{cond1}) and (\ref{cond2}) are satisfied for
these $\zeta$ and $\nu$ if
\begin{equation}
 \tau < \tau_{\ast} = {2 \over 9} \, \left( 1 + {\psi \over {1 +\gamma -\psi}}
\right) \ .
\end{equation}
	In the $\psi \rightarrow 0$ limit $\tau_{\ast}$ corresponds to the temperature
of the first-order phase transition $T_{\ast}$ in the absence of stresses. An
important difference is, however, that now the product phase does not
correspond to the free energy minimum above this temperature and, therefore, is
{\em mechanically} unstable. The energy dependence on $\zeta$ is shown in
Fig.\ref{EnergyZeta} for $\tau < \tau_{\ast}$. There is an activation barrier
separating the initial $\zeta = 0$ state and the product phase that corresponds
to minimum of free energy for $\zeta \ne 0$. However, there is no barrier in
the free energy dependence on $\nu$ shown in Fig.\ref{EnergyNu}. It means that
there exists a transformation path in the $(\zeta,\nu)$ phase space that does
not involve any activation process.

\begin{figure}
\epsfxsize 5.75in
\centerline{\epsffile{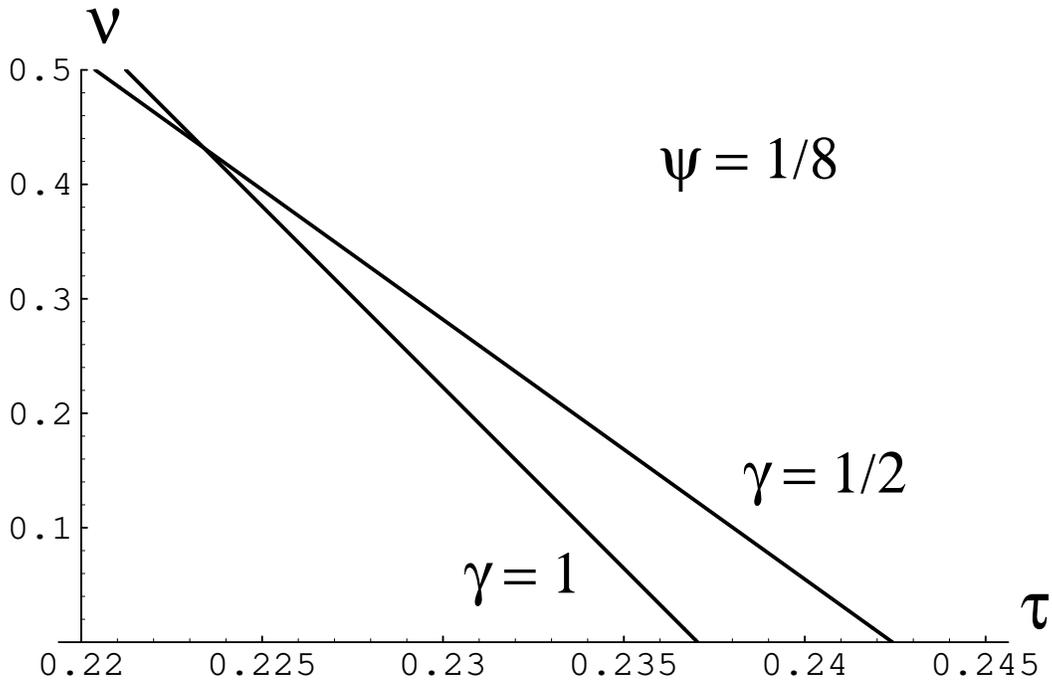}}
\caption{Temperature dependence of the equilibrium volume fraction of the new
phase $\nu$ for two different values of the ratio of elastic moduli $\gamma$.}
        \label{Nutau}
\end{figure}

The classical phase rule allows the equilibrium co-existence of the parent and
product phases of the same composition only at the fixed temperature
$T_{\ast}$. However, the phase rule is based on the assumption that there is no
elastic interaction between the phases. The finite equilibrium volume fraction
of the second phase $\nu$ that corresponds to the minimal $\Delta \tilde{\cal
G}$ in the present model appears in some temperature interval below
$\tau_{\ast}$ and depends linearly on the dimensionless temperature $\tau$
according to Eq.(\ref{nutau}) which is illustrated in Fig.\ref{Nutau}. The
system has to be cooled further down for the increase in $\nu$, i.e. for the
transformation to proceed.

The present model corresponds to the `athermal' kind of martensitic
transformation in metals when the transformation begins at some start
temperature $M_s$, but the parent phase still exists until the temperature goes
down to $M_f$, a martensite finish point. The finish point $M_f$ corresponds to
$\nu = 1$ and the present model can not be used in this region because the
inclusions of the new phase are supposed to be well separated.

In the absence of stresses the transformation rate is determined by Arrenius
factor for the activation energy barrier as well as by the nucleation and
growth kinetics. The system can be supercooled and the transformation proceeds
with a finite rate at any constant temperature below $T_{\ast}$ and can be
completed at the same temperature in a finite time. In the present model
the transformation does not involve any (thermal) activation process as there
is no energy barrier. It means that fluctuations do not play an important role,
supercooling is impossible and there is no Arrenius factor in the
transformation rate which therefore should be very high. Such a `non-activated'
kinetics is a characteristic feature of the martensitic transformations. If the
temperature interval between $M_s$ and $M_f$ is narrow enough than for any real
cooling rate transformation would be completed almost instantly. This is known
to be the case for so-called `isothermal' kind of martensitic transformations
in metallic alloys.

To conclude we have analyzed an effect of elastic stresses in the matrix on the
thermodynamics and kinetics of the first-order phase transformation in an
embedded inclusion. The coupling of strain with the transformation degrees of
freedom has been considered in the frame of Landau theory of phase transitions.
The transition takes place in some temperature interval rather then at a
certain temperature and the transformation kinetics appears to be
activationless. This model may describe martensitic transformations in some
metallic alloys.

\acknowledgments

Discussions with A. Artemev, V. Breiguine and A.R. Roytburd were very helpful.
The work would have not been done without hospitality of Prof. J. Goldak at
Carleton University.

\end{document}